\documentclass[11pt]{article}                  
  \usepackage{times}                   

 \usepackage[dvips]{graphicx}

\begin{document}
\title{Scaling and singularity characteristics of solar wind and magnetospheric fluctuations}

\author{Z. V\"{o}r\"{o}s (1), 
D. Jankovi\v{c}ov\'{a} (1),
P. Kov\'{a}cs (2) \\
(1) Geomagnetic Observatory Hurbanovo, Geophysical Institute SAS, 
Slovak Republic,\\
(2) E\"{o}tv\"{o}s Lor\'{a}nd Geophysical Institute, Hungary,
kovacs@elgi.hu \\
}
\date{}
\maketitle

\begin{abstract}
Preliminary results are presented which suggest that scaling and singularity characteristics
of solar wind and ground based magnetic fluctuations appear to be a significant 
component in the solar wind - magnetosphere interaction processes. 
Of key importance is the intermittence of the "magnetic turbulence" as seen in ground
based and solar wind magnetic data. 
The methods used in this paper (estimation of flatness and multifractal spectra) 
are commonly used in the studies of fluid or MHD turbulence.
The results show that single observatory 
characteristics of magnetic fluctuations 
are different from those of the multi-observatory AE-index. 
In both data sets, however, the influence of the solar wind fluctuations is 
recognizable.
The correlation between the scaling/singularity features of solar wind magnetic 
fluctuations and  the corresponding geomagnetic response is demonstrated in a 
number of cases. The results are also discussed in terms of patchy reconnection
processes in magnetopause and forced or/and self-organized criticality (F/SOC) of
internal magnetosphere dynamics.
\end{abstract}

\section{Introduction}\label{sec:intro}

According to recent knowledge magnetospheric plasma fluctuations may be explained 
in terms of multiscale intermittent  models including nonlinear self-organization 
processes (forced or not) near criticality  (F/SOC) 
\cite{chan92, chan99, chap98, taka99, klim00}. 
The basic physical concepts and the corresponding path-integral and renormalization 
group formalisms of nonlinear self-organization of magnetospheric processes 
are described in \cite{chan99}.
These models predict multifractality, non-Gaussian, power-law distributions
for certain measurable physical quantities and  associated power-law scalings  
for power spectra. In fact, several experimental studies seem to support the theoretical
predictions quite well 
\cite{cons96, milo96, cons97, boro97, cons98,voro98, urit98, conl99, chap99, frem00, frea00, sha01}. 
Scale-free power spectra of magnetospheric fluctuations, however, admit several 
possible explanations other than F/SOC. \cite{watk01} compared linear noisy, 
low-dimensional nonlinear, stochastic fractional Brownian motion (fBm) and SOC 
models of magnetospheric 
fluctuations to examine their scaling and predictabilitity properties. 
They concluded that the SOC model is of particular interest to magnetospheric 
physics due to its robustness in 
explaining scalings under the wide range of activity levels exhibited by 
the magnetosphere and the solar wind. On the other hand, weakly nonlinear 
models with fBm type noise added could explain the scaling and predictabitity 
properties of magnetospheric fluctuations equally well.
Geomagnetic fluctuations on the time 
scale of substorms and storms certainly appear to be a compound mixture  of 
multiscale magnetospheric processes including the component represented by  
intermittent 
solar wind fluctuations \cite{chan99, dagl99}. Spectral methods making a 
distinction between  these 
constituents  may be queried to a certain extent mainly because the information on 
nonlinear multiscale structures is partially hidden for second order statistics. 
For example, \cite{tsur90} using the AE-index time series have shown that 
the power spectrum of the AE data exhibits two different power law scalings divided 
by a 
spectral break at the frequency range about 1/(5h). While the higher frequency 
part  was 
thought to be more intrinsic to the magnetosphere, the lower frequency part 
was attributed 
to the  influence of solar wind. \cite{voro00} has shown that 
similar results can be 
obtained also by means of a multifractal technique,  
which also makes possible, however, 
to reveal additional information on singularity distributions 
of high-latitude geomagnetic fluctuations (see later), 
not evident from spectral studies. Among other things, the results of multifractal 
analysis
 show that the local singularity (H\"{o}lder) exponents are time dependent 
and the 
previously proposed fBm type or bicolored noise model \cite{taka93} of 
geomagnetic fluctuations is
 not relevant. Moreover, for a range of singularity 
exponents, 
a multiplicative cascade model, (the $P$-model) describes quite well 
the observed 
singularity distribution of geomagnetic fluctuations. Out of this range,
however, significant deviations from a multiplicative model appear.
The $P$-model describes energy cascade processes in turbulent flows. 
The largest turbulent eddy is assumed to be built up by a specific energy flux per 
unit length. Then a scale-independent space-averaged cascade-rate is considered 
and the flux density is transferred to the two smaller eddies with the same length 
but different flux probabilities $p_{1}$ and $p_{2}$. This process with randomly 
distributed $p_{1}$ and $p_{2}$ ($p_{1}$ + $p_{2}$ = 1) is repeated towards smaller 
and smaller scales. Energy transfer rate is homogeneous for $p$ = $p_{1}$ = $p_{2}$ = 0.5 
while $p_{1}$ $>$ 0.5 corresponds to an intermittent flow. 
Figure 1 shows how the intermittence increases with increasing parameter $p_{1}$ 
\cite{tuuu96} in  $\alpha$, $f(\alpha)$ plane \cite{hal86}. The larger spread of 
$\alpha$ values around right-shifted $\bar \alpha$, the more intermittent field. 
In spite of the simplicity of 
the outlined cascade model its relevance for the description of intermittence 
effects in magnetospheric fluctuations may lead to the assumption that turbulence 
rather than F/SOC models fit the observed statistics better. Some results on 
statistical distribution of internal time periods 
between bursty geomagnetic events (waiting times) \cite{kovz01} seem to support this
assumption. SOC models (at least the original \cite{bak87} model) are expected to display 
an exponential waiting time distribution. Geomagnetic data, however, display a 
well-defined power-law waiting time distribution.
It was pointed out by \cite{boff99} that the observed power-law waiting time statistics 
in solar flares appears to be well explained by MHD shell models of turbulence.
Similar results were presented by \cite{spad01} who analysed density fluctuations in
a magnetically confined plasma system and found that waiting time statistics is 
in contrast with the predictions of an SOC system.
An opposite view was presented by \cite{frea00} who conjectured that a wider class of 
running sandpile models \cite{hwa92} could exhibit power law behaviour in the probability 
density functions of waiting times. \cite{watm01} determined that the PDFs for burst
durations and waiting times in a reduced MHD simulation follow power-laws
which is not sufficient to distinguish between turbulence, SOC-like models and
colored noise sources.
\cite{boff99} and \cite{ant01} also argued that SOC models represent self-similar, 
fractal phenomena. 
Geomagnetic fluctuations exhibit clear multifractal scaling \cite{cons96, voro00} 
which seems
to contradict to SOC concepts again. \cite{geor95}, however, demonstrated that the SOC 
state displayed by their cellular automaton model of isotropic and anisotropic 
energy avalanches has multifractal and multiscaling characteristics, 
rather than single power-law scalings and this feature was even enhanced 
by considering extended instability criteria. Similar SOC models of solar flares 
also exhibit multifractal and multiscaling characteristics \cite{vlah95}.
Recent works \cite{vasi98, isl98, isl00, geor01, urit01a} 
underline the proximity between the SOC rules and laws of MHD 
in space physics systems, which are long known to exhibit turbulent behaviour 
(Georgoulis, personal communication, 2001). More realistic F/SOC models 
which include e.g. less artificial 
feedback mechanisms, a wide variety of drivings, interacting avalanches, 
SOC in continuum physical systems \cite{lu95}, etc. may further establish 
the proximity with turbulence.
As pointed out by \cite{urit01a,b} the concept of SOC in a continuum limit 
(fluid or MHD limit) is essentially unexplored. \cite{klim00} proposed a simplified 
Earth's magnetotail current sheet model based on continuum SOC model of \cite{lu95}.
The continuum Lu model in a magnetic field reversal configuration can evolve 
into SOC due to localized rapid magnetic field annihilation within the field 
reversal region. In the same time, the plasma sheet is dominated by strong 
turbulence which keeps the system near criticality and produces a predictable 
quasi-periodic loading-unloading cycle of coherent global substorm activity 
\cite{klim00}. In this model turbulence, SOC states and coherent global modes
coexist within the Earth's magnetotail on different scales. 
As a result, the observed ground based and satellite time-series contain a mix 
of fluctuations of different physical origin. \cite{ange99} further argued that 
the presence of intermittent turbulence in the Earth's magnetotail may alter 
the conductivity and the mass/momentum diffusion properties across the plasma 
sheet and may permit cross-scale coupling processes playing also an important 
role in the establishment of SOC state.  

In this paper no attempt will be made to participate in the theoretical debate 
on SOC or turbulence. However, having a pragmatic view, our opinion is that 
measures of intermittence
or characteristic descriptors of cascade processes, commonly used in 
turbulence studies, like the flatness and multifractal spectra, could be applied 
in solar wind-magnetosphere interaction studies for further comparison of basic 
characteristics of intermittent fluctuations in solar wind and within the 
magnetosphere. We assume this aproach might be useful providing experimental 
information on such characteristics of fluctuations which are not accessible for 
spectral studies or second order statistics.
This assumption was already investigated in \cite{voro98, voro00, voro01, kovz01}.
The preliminary results allow to make a working hypothesis
that intermittence, scaling, rapid changes, singularities represent an essential
piece of information regarding the effectiveness of SW - magnetosphere coupling not
considered enough hitherto.
In order to go deeper, we analyse different geomagnetic and solar wind data sets and make 
a comparison between various
 magnetic activity levels considering time scales 
of geomagnetic storms 
(from  hours to days), substorms (from half an hour 
to a few hours) or less. 

The main goal of this comparative
study is to contribute to the understanding of solar wind - magnetosphere 
interaction
 processes on the basis of characteristic scaling/singularity features 
of the considered
 time series. We recall
 that solar wind (SW) fluctuations are
strongly intermittent \cite{burl91}, that is energy at a given scale is 
not homogeneously
 distributed in space or/and time. Several studies on 
characteristic probability
 distribution functions (PDF) of increments of SW 
parameters (magnetic field, velocity,
 temperature, Els\"{a}sser variable, etc.) 
\cite{mars94, sorr99}
 and on multifractal structure of SW fluctuations 
\cite{burl92, carb94, mars96, tuuu96}
 support this assumption. Moreover, 
\cite{vel99} and \cite{brun99} have shown that there 
is a direct link between intermittence 
and the presence of SW structures (Alfv\'{e}nic, 
magnetic fluctuations, 
discontinuities) across which the magnetic field magnitude
 changes. Also, 
the high frequency (small scale) fluctuations of the southward component
of the interplanetary magnetic field have a different spectral scaling 
exponent as the
 one exhibited by geomagnetic AE index fluctuations. 
On larger scales the corresponding
 spectra are similar \cite{tsur90}.
A comparative study of dynamical critical scalings in the auroral electrojet 
(AE) index versus solar wind fluctuations confirmed that for times shorter 
than 3.5 hours (higher frequencies) the AE index fluctuations are of internal 
magnetospheric origin \cite{urit01a}.
 
\section{Data analysis methods}
\label{sec:mul}

As usual, we introduce the concept of scale ($\tau$) through the difference
\begin{equation}
\delta X(t,\tau) = X(t+\tau) - X(t) 
\end{equation}
where $X(t)$ is the time series under consideration. \cite{mars94} have shown
that the PDFs of the increments $\delta X$ ($X$ - SW parameters) exhibit strong
deviations from Gaussianity, especially at smaller scales and the effect is due to 
intermittence of SW fluctuations. To quantify the degree of deviation from 
the Gaussian
 distribution, i.e. the level of intermittence at different scales, 
we compute the flatness
 defined by
\begin{equation}
F = \frac{<\delta X(t,\tau)^4>}{<\delta X(t,\tau)^2>^2}
\end{equation}
The flatness of a normally distributed signal is equal 3. Adding intermittent 
fluctuations
 to an originally Gaussian signal implies the spreading of its PDF, 
and consequently, the 
increase of its flatness from the original value of 3. 

Non-homogeneous/intermittent distributions in space/time  may also appear as 
asymptotically
 singular and can be characterized locally, at the point $t_{i}$, 
by the singularity (H\"{o}lder) exponents 
$\alpha$ as 
\cite{vehe96, vehe98, ried95, canu98}

\begin{equation}
\alpha_{n}^{k}(t_{i}) = \lim_{n \rightarrow \infty } \frac{-log \mu (I_{n}^{k}(t_{i}))}{n}
\end{equation}
where $\mu$ is a measure constructed from a time series. The procedure involvs the 
computation of the energy content of 
the differenced signal (Eq. 1) by taking its squared value. 
The measure at a point $t_{i}$ is given by 
$\delta X ^{2}(t_i,\tau) / \sum_{i} \delta X^{2}(t_i,\tau)$.
To analyse the distribution of singularity exponents (Eq. 3)
a sequence of partitions $P$ is
introduced so that \cite{vehe96}
\[ P_n = \{I_{n}^{k}\} ;  \:\:\:\:\:\:\:\: 0 \le k < 2^{n}-1 \]
\[ I_{n}^{k} = <k2^{-n},(k+1)2^{-n}) \]
where $I_{n}^{k}$ is the interval containing t and the resolution is set by $n$. 
The quantity 
of interest is the so-called large deviation singularity spectrum, 
$f(\alpha)$, which
 represents a rate function measuring the deviation of the 
observed $\alpha$ from the
 expected value $\overline {\alpha}$. The rate function, 
$f(\alpha)$, can be estimated 
through \cite{vehe96}
:

\begin{equation}
f(\alpha) = \lim_{n \rightarrow \infty}\frac{N_{n}(\alpha)}{n} 
\end{equation}
where $N_{n}(\alpha)$ is the observed number of coarse grain H\"{o}lder 
exponents
. Usually "histogram 
methods" for 
estimation of $f(\alpha) $ are used. In that case
the number of those intervals $I_n^k$ for which $\alpha _n^k$ falls in a box between
$\alpha _{min}$ and $\alpha _{max}$ is computed and $f(\alpha) $ is found by 
a regression.
 It yields satisfactory results for pure multiplicative
processes, but fails to describe non pure or compound processes when $f(\alpha)$  is
not a concave function.  To overcome this difficulty the so-called 
double kernel method was proposed \cite{vehe96, vehe98}
realizing that $N_n(\alpha)$ may be written
as a convolution of the density of the $\alpha_n^k$s and a compactly supported 
kernel. 
This method allows to estimate non concave rate functions and we are 
going to show that 
this property may be properly used for characterization of 
fluctuation processes in 
near-Earth space. In this paper the estimations 
of $f(\alpha)$ spectra were realized using
 the FRACLAB package developed 
at the Institut National de Recherche en Informatique, 
Le Chesnay, France.

\section{Ground based data}
\label{sec:gro}
In order to study the basic characteristics of auroral zone 
geomagnetic fluctuations
 we analyse geomagnetic 
$H$-component 1-min mean data from polar cap observatory, 
THULE (THL: $77.47^{o}$ N, $290.77^{o}$ E),
geomagnetic $X$-component 1-min mean data from high-latitude observatory NARSSARSSUAQ 
(NAQ: $61.16^{o}$ N, $314.57^{o}$ E) and 1-min mean auroral electrojet (AE) index 
time series, all from  1991-1992.

As known, the 
AE-index was introduced by \cite{davi66} to describe the global activity of 
the auroral zone electric currents and is derived, after the substraction of 
base line
 values, from evaluation of the variations measured at 12 stations 
located near the 
northern auroral zone. There exist a large number of physical 
mechanisms which couple the 
auroral zone processes with those within 
magnetospheric tail or in SW. Recently, intermittent energy transport 
in the magnetotail, the so called bursty bulk flow events (BBF)
 came into 
the limelight of the magnetosphere research \cite{baum90, ange92}. 
From this point of view the
 understanding of the response of auroral zone 
currents or dissipation fields to the 
time - varying magnetotail dynamics 
seems to be important. There was not full understanding
 achieved regarding 
the nature and origin of the related magnetic fluctuations. Partly it was 
already explained in the introduction that it is related to the  
paradigms of F/SOC versus turbulence. We mention here some other open questions. 
For example it was shown that the burst lifetime distributions of some SW 
parameters are also of power law form, which
 might be a signature of SOC 
or turbulence regimes in SW \cite{frem00, kova01}. Therefore, it is supposed 
that the scale free
 property of the AE-index may arise from the SW input 
or at least the internal dynamics
 of the magnetosphere may be masked by the 
scale free properties of SW driver \cite{frem00}. 
Again we remind, however, the very limitations of second order statistics 
in interpretation of the observed scalings. 
Another
 measure of the auroral zone dissipation fields is represented by 
polar optical activity
 within UVI bands. \cite{luii00} have examined the blobs 
of brightness as a proxy
 for BBF events. It was found that the non-substorm 
"internal" events have a power law
 distribution whereas the system wide events 
like substorms besides a scale free region
 exhibit  a "bump", corresponding 
to a mean value in substorm breakups.  
A somewhat opposite 
view was presented by \cite{cons98} who analysed AE-index 
fluctuations
 on time scales 1-120 [min] both in quiet (laminar phase) and 
disturbed periods (turbulent
 phase). They found that in both phases the 
intermittence at different time scales rescales
 in the same way and the 
non-Gaussian character of PDFs seems to be due to the same physical
processes.   

Here we pose the question again about the scaling and singularity properties of the 
AE-index, compared to the similar characteristics of geomagnetic data from two
observatories THL and NAQ. As the AE index is derived from geomagnetic variations 
in the horizontal 
component observed at selected  observatories along the auroral 
zone in the 
northern hemisphere, we expect that the scale $\tau$ (Eq. 1) cannot 
be precisely
 defined. The geographic distance between observatories through 
Taylor's hypothesis already
 introduces some effective time shift (scale, $\tau$). 
Though, the application of
 the Taylor hypothesis within the magnetosphere 
is limited \cite{dudo96}. Besides, from the 
recordings
of  auroral stations the greatest (upper envelope) and smallest (lower envelope)
values are taken at intervals of one minute and their difference defines the 
AE-index.
As far as the contributing observatory which gives the lower/upper envelope changes
during the times, and the fluctuations with values between the upper and lower 
envelopes are not taken into account at all (smoothing), the AE-index appears to 
be a measure
 of auroral zone processes with mixed scales. This is certainly not 
an advantage when a
 multiscale analysis of AE-index time series is performed. 
This fact is usually neglected
 in the related literature. A less sophisticated  
way is to take data from a single 
observatory, but it may have some other 
drawbacks because significant distant 
disturbance events  can be missed. Also, 
some BBFs with short duration may remain undetected 
on the ground because of 
their localized nature \cite{dagl99}. 
Nevertheless, a comparison of fluctuations of the 
"multi-observatory measure" 
(MOM: AE-index) and of  the "single observatory measure" 
(SOM: THL geomagnetic 
field $H$-component and NAQ geomagnetic field $X$-component) may be instructive.

Figure 2 shows the PDFs from normalized increments $\delta X$ (Eq.1) of the SOM 
(Figure 1a, b)
 and MOM (Figure 1b) data (X $\equiv$ X(NAQ), H(THL) and AE, 
respectively).
Calculations have been made for $\tau$ = 5, 50, 500, 5000 [min] 
and 1 minute mean data was considered from the years 1991-1992. As can be seen the
distributions in all cases change with $\tau$, and for smaller values of $\tau$ 
significant  deviations from the normal distribution occur.
The tails of the distributions reduce with increasing scale parameter as a 
consequence
 of the decrease of the probability of coherent fluctuations between 
points separated by 
increasing distance. The difference between the MOM and SOM 
is more clear if the flatness
 ($F$) of the corresponding distributions is compared 
(Eq. 2). Figure 3 shows how the
 flatness evolves with increasing $\tau$ 
($\tau$ $\in$ (5, 5000) [min]).
The errorbars correspond to the standard deviations (std) computed from time
series divided to several parts. For small scales,
 say, $\tau$ less than a few 
tens of minutes, the AE-index (MOM) exhibits smaller deviations from the 
normal distribution than NAQ (SOM). $F(NAQ)$ reaches the level of 
$F_{max}$ (AE, $\tau$ = 5) only
at the value $\tau$ $\sim$ 30 [min], which roughly may be considered as the effective
time shift introduced by the method of derivation of the AE-index. 
THL (polar cap observatory) data essentially show the same behaviour as NAQ, but the 
stds are larger.
Therefore, we conjecture
 that MOMs (multi-observatory geomagnetic indices), 
due to smoothing and scale mixing
 effects, lead to underestimation of the 
intermittence on small scales. 
Henceforth, we will show only the dependence 
of the flatness on scale parameter $\tau$.

Let us consider now the singularity distributions $f(\alpha)$ (Eq. 3, 4) for the same
data sets as above. To see better how the $f(\alpha)$ rate functions evolve with
$\tau$, all the curves are depicted at the same plane $(\alpha, f(\alpha))$ 
in Figure 4.
Proper symbols are introduced for the scales $\tau$ = 10, 50, 500 [min] 
($\bigtriangleup, *, o$) and the dashed lines correspond to the
errors estimated by changing the resolution in Eq. 3, 4. (No averaging needs 
to be done
 estimating $f(\alpha)$ \cite{vehe96} so the spectrum may be 
evaluated at only
 one resolution. However, to show that the estimations are 
consistent, all the spectra 
were computed using 15 different resolutions). 
Again, there are several differencies
 between the SOM and MOM data.
Figure 4a shows the singularity spectra for NAQ observatory data. On the considered 
scales, the shape of the curves is almost parabolic, close to the $P$-model fit 
(thick curve) with  $p_{1} \sim 0.745$. The best correspondence is achieved for the 
smallest value of $\tau$ = 10 [min] (depicted by symbol $\bigtriangleup$ 
on Figure 4a). It means that in case of auroral zone SOM fluctuations and 
especially on small scales, the deviations from the Gaussian distribution 
(Figure 3) can be explained by a simple cascade model. 
Though the phenomenology of turbulent cascades in fluid flows is more complex than 
the simple $P$-model fit in Figure 4a would lead us to indicate, the 1D cascade 
model rouhgly describes how the auroral zone SOM fluctuations become more and more 
intermittent at smaller and smaller scales. The spectra for polar cap 
(THL observatory) SOM (Figure 4b) and AE index MOM (Figure 4c) have a more 
pronounced non-parabolic shape indicating the presence of compound processes. 
At small scales ($\tau$ = 5,50 [min]) THL observatory fluctuations contain 
stronger singularities because of the extension of the rate functions left 
wings to the smaller (more singular) values of $\alpha$. For $\tau$ = 500 [min], 
however, the right wing evolves to the less singular values which may be related 
to the SW influence \cite{voro00}. This effect is less visible, but still present 
in auroral zone SOM data (Figure 4a). The MOM AE index $f(\alpha)$ spectra 
practically do not change with $\tau$. We conjecture, this is the result 
of the method of derivation of the AE index, resulting smoothing and scale 
mixing effects. 

In all cases, the deviations of singularity spectra from 
the parabolic shape may be indicative of the phenomenon of phase transition.
Namely, at the $\alpha$ values where the $f(\alpha)$ spectra are out of 
parabolic shape the
 major contributor to the observed singularities may change 
from one measure to another.
As different physical processes may generate different measures (distributions), 
possible 
models with similar characteristics as the observed spectra may contain 
physical 
information on the contributing (e.g. SW or magnetospheric) sources.

\section{Satellite data}
\label{sec:sat}

The very advantage of the ground based data is its availability for a long periods 
of time.
 For a proper estimation of PDFs or singularity rate functions  long data 
sets are
 needed which is a requirement hardly ever fulfilled  in the case of 
satellite data. 
Nevertheless, we expect to find out some interesting 
scaling/singularity
 features of interplanetary magnetic field (IMF) fluctuations 
proceeding in the same way
 as in the previous section. To this end we analyse 
ACE and WIND IMF $B$ magnitude and $B_{z}$ component data
which are available with time resolution of 16 [s] and 3 [s], respectively. 
SW velocity is not considered here due to too many gaps in data.
While the ACE
 satellite is continuously monitoring the SW at the $L_{1}$ point, 
WIND has a more complicated 
trajectory crossing also the magnetosphere from time 
to time. For our analysis we have chosen
 time periods when WIND was also in SW 
and there were  negligible data gaps in both cases
 (less than 1 $\%$ of total 
data lengths). 

Another aim was to analyse "geoeffectively 
different" periods 
of IMF $B$ and $B_{z}$ magnitude fluctuations. 
Geoeffectiveness during the chosen
 periods was considered examining the 
geomagnetic $D_{st}$ index which is derived
 from the geomagnetic field $H$-component 
registrations of 4 observatories \cite{sugi64}
 and it aims at giving the effect 
of the magnetospheric ring currents.
 The chosen periods were classified as 
disturbed ones if geomagnetic fluctuations with storm-index $D_{st}$ being less 
than -50 [nT]
 occured several times within a consedered interval.
The limit of $D_{st} <$ -100 [nT] which corresponds to intense storms was 
considered, too. 
We emphasize, however, instead of a study of individual storms, 
the generic features of fluctuations on a given scale $\tau$, but 
during longer periods of time are investigated. Essentially 2 $\div$ 4 
weeks of data with the consedered time resolutions may already ensure sufficiently 
robust estimations of singularity spectra (Eq. 4). In this sense, several intense 
magnetic storms $D_{st}$ $\ll$ -50 [nT] may occur during a strongly disturbed 
interval, a less disturbed period contains less intense storms and an undisturbed 
period has only $D_{st} >$ -50 [nT]. The limit of -50 [nT] was chosen on the basis 
of previous studies of magnetic storms \cite{tay96}. Intense magnetic storms are 
characterised by $D_{st}$ index $<$ -100 [nT] \cite{gonz87}. In this preliminary study 
of generic features of magnetic fluctuations these limits may be consedered as more 
or less adequate. We expect that this rough classification of the geomagnetic 
response allows us to identify characteristic scaling and singularity features of 
the corresponding IMF magnetic fluctuations that would be indicative for their 
geoeffectiveness.  

There were 5 time periods and 6 data sets separated for our analysis (for one 
period there
 were both ACE and WIND data available). The disturbed periods are 
the 
following: $1.$ March 19 - April 25, 2001 (ACE); $2.$ October 1 - November 30, 
2000 (ACE);
 $3.$ April 9 - April 20, 1997 (WIND). The undisturbed periods: $4.$ 
November 18 - December 10, 
1998 (ACE); $5.$ January 10 - January 29, 1998 
(ACE, WIND).
 For demonstration we show some of the data sets.

Figure 5 shows the first ACE data set from March 19 to April 25, 2001. For the 
sake of perspicuity, the time resolution
 is 1 [h]. 
(The flatness and singularity spectra are computed from the time series 
with time resolution of 16 [s] and 3 [s].)
During this extremely active period intense magnetic storms ($D_{st}$ $\leq$ 
-100 [nT]) occured several times (Figure 5c). The limit of -100 [nT] is 
depicted by a thick line in Figure 5c. \cite{gonz87} have shown that the 
interplanetary causes of intense magnetic storms are long duration ($>$ 3 [h]), 
large and negative ($<$ -10 [nT]) IMF $B_{z}$ events associated with 
interplanetary duskward electric fields $>$ 5 [$mVm^{-1}$]. In Figure 5b IMF 
$B_{z}$ is depicted, including a thick line indicating the level of -10 [nT]. 
Comparison of Figures 5b, c shows an agreement with the above criteria, that is, 
long duration negative IMF $B_{z}$ events occur together with intense magnetic 
storms. Figure 5a shows the variations of IMF $B$. It is visible that an 
intense magnetic storm occured at the end of the studied period, 
between $t$ = 800 and 900 [h] (Figure 5c). $B_{z} <$ 10 [nT] (Figure 5b) and 
$B_{max} \sim$ 17 [nT] corresponded to this event. A similar enhancement of $B$ 
at 400 [h] $\leq$ $t$ $<$ 450 [h] appears in Figure 5a having no intense 
storm response in $D_{st}$, which can be explained by the corresponding IMF 
$B_{z} >$ -10 [nT] in Figure 5b.

Figure 6 shows an undisturbed period from January 10 to January 29, 1998. Again, 
IMF $B$, $B_{z}$ (WIND, ACE Figures 6a-d) and $D_{st}$ index (Figure 6e) are shown. 
It is visible that $D_{st} >$ -50 [nT] and $B_{z} >$ -10 [nT] everywhere.

Fluctuations of IMF and their geoeffectiveness were studied by a number of authors. 
\cite{mcp86}
showed that substorms are frequently triggered by changes in the IMF. 
\cite{kami01} proposed that the quasi-steady component of the interplanetary 
electric field is imporant in enhancing the ring current, while its 
fluctuations are responsible for initiating magneospheric substorms. 
It is out of scope of this paper to analyse the influence of other interplanetary 
parameters (e.g. velocity, density, temperature, etc.) on storm/ substorm activity
\cite{dagl01}. 
Rather we will concentrate on the level of intermittence of IMF fluctuations.

To this end let us consider the flatnesses and the singularity rate functions 
for the disturbed
 and undisturbed periods  of $1 - 5$. Deviations from the 
Gaussian distribution are 
larger in the case of disturbed events depicted by 
larger markersizes in Figure 7. The 
PDFs of the undisturbed events are also 
non-Gaussian, but the flatnesses for a given
 scale are smaller than those for 
disturbed events especially at scales $\tau <$ 1000 
[s]. Similar differences 
are present in singularity spectra computed for IMF $B$ fluctuations at the 
scales of
 $\tau$ = 320 [sec] (ACE) and $\tau$ = 60 [s] (WIND) shown in Figure 8. 
There were the same marker types used as in Figure 7.
 ACE and WIND data are 
depicted separately in Figures 8a, b.
 The maxima of the singularity spectra 
of more disturbed periods have a tendency
 for shifting  to larger values of 
$\alpha$. Also, the spread of singularities around most probable $\alpha$ 
is wider for more disturbed cases. 
But it is the same behaviour as in case of the simple $P$-model in Figure 1, when 
intermittence is stronger and stronger for larger and larger values of $p_{1}$.
The differences between the disturbed and undisturbed cases gradually cease 
for larger values
 of $\tau$ (not shown).
 
Figure 9 shows the singularity spectra computed for IMF $B_{z}$ fluctuations 
in the same way as previously. Obviously, the intermittent fluctuations of IMF 
$B$ and $B_{z}$ fields exhibit very similar changes in their flatnesses and 
singularity spectra as the geoeffectivity level changes. For example, there is 
a clear difference between the introduced scaling and singularity characteristics 
(Figure 5) for the disturbed period March 19 - April 25, 2001 and for the 
undisturbed period January 10 - January 29, 1998 (Figure 6).
It indicates that, in addition to known "geoeffective"  SW parameters 
(e.g. southward component of IMF) or their combinations, small scale rapid 
changes, 
singularities  and non-Gaussian statistics of IMF fluctuations may play 
an important role 
in SW - magnetosphere interaction processes. 

Also, the question naturally arises to what extent the magnetospheric response 
itself is 
influenced  by small scale statistics of IMF fluctuations. 
Unfortunately, not all the geomagnetic data is available for the above analysed 
periods. It is possible to test,
 however, how the shape of the rate function 
changes if available data is 
considered. 
Minute-mean $H$-component geomagnetic data from THL observatory is available from 
1975 to 1996. 
We computed the $f(\alpha)$ spectra for each year at the scale $\tau$ = 50 [min]
and analysed how their shapes change at the $f(\alpha)$ values 0.3 - 0.9.
At each level of 
$f(\alpha)$ the corresponding values of $\alpha_{min}$ and $\alpha_{max}$
were computed (filled circles in Figure 10a). In Figure 10b the time evolution of the 
difference $\alpha_{max} - \alpha_{min}$ at a given $f(\alpha)$ level is depicted.
The average standard deviation at $f(\alpha)$ = 0.3 is about 0.06, while at 
$f(\alpha)$ = 0.9 is 0.02.
At $f(\alpha)$ = 0.3 - 0.6 the $\alpha_{max}-\alpha_{min}$ curves strongly fluctuate 
indicating significant changes in the shape of $f(\alpha)$ spectra from 1975 to 1996. 
Figure 11 shows the similar results of \cite{kami98}, however, obtained by a different 
method. \cite{kami98} analysed the occurence of geomagnetic storms in comparison with 
yearly averaged Wolf sunspot number. Figure 11 shows the yearly averaged number 
of hours with $D_{st}$ less than -100 [nT] (solid line with filled diamonds), and 
with $D_{st}$ less than -50 [nT] (divided by 5, dashed line). Thick line corresponds 
to yearly averaged sunspot number. It is visible that the maxima of geomagnetic 
activity and of solar cycle do not coincide. During the declining phase of the 
solar cycle coronal holes emerge from polar regions of the Sun which are continuous 
sources of fast-speed plasma causing a peak in recurrent geomagnetic storms 
activity \cite{kami98, kami01}. The similarity between the variability of rate function shapes 
for $f(\alpha)$ = 0.3 $\div$ 0.6 (Figure 10b) and yearly averaged number of hours 
with prescribed $D_{st}$ indices (Figure 11) is remarkable. This correspondence 
also supports our working assumption that the shape of $f(\alpha)$ rate function 
estimated using Eqs. 3, 4 \cite{vehe96, vehe98} contains relevant physical information. 
We mention that available AE-index and NAQ observatory data lead to the same 
results. On the other hand, however, at $f(\alpha)$ = 0.7 - 0.9 the variations of 
$\alpha_{max}-\alpha_{min}$ versus time are negligible (Figure 10b). 

On the basis of Figures 10b and 11, years with extreme levels of geomagnetic 
response can be chosen and their corresponding singularity spectra can be 
recalculated. Figures 12a-c show the SOM 
and MOM spectra estimated at $\tau$ = 50 [min] for maximum (1991, symbol o) 
and minimum (1984 - symbol x) years of geomagnetic activity. One can see that 
the characteristic asymmetric shape of AE-index spectra (Figure 12c) in 
comparison with Figure 4c is present henceforward, presumably caused by 
the influence of SW fluctuations. Noticeably, the maxima and the right 
wings of the more geoeffective SW IMF singularity distributions (Figures 8, 9) 
match well the region of $\alpha \in$ (1.2, 1.8) within which the MOM spectra 
for the most part are out of parabolic shapes. The same effect is visible for 
the more singular wing of $f(\alpha)$ spectra at $\alpha$ $\in$ (0.4, 0.8). 
The asymmetry is even enhanced in 1991 (maximum of geomagnetic activity). 
The differences between more active (1991) and less active (1984) years are 
present in SOM spectra mainly at the wings of $f(\alpha)$ rate function, too 
(Figures 12a, b). The most probable singularities around $\bar \alpha$ do not 
exhibit any changes.

\section{Discussion}
\label{sec:dis}

We presented an analysis of scaling and singularity characteristics of 
ground based and satellite magnetic fluctuations
in this paper. 
These techniques are commonly used in studies of turbulent flows. 
As we oulined in the Introduction the supposed proximity between F/SOC 
and MHD turbulence models may allow to estimate measures of  
scaling, intermittence, etc.,  directly from time series, for futrher comparison 
aiming to help a development of tractable numerical models of highly 
variable SW - magnetosphere interaction 
(Georgoulis, personal communication, 2001).
We think the results are not contradictory in this sense, but rather 
elucidate important aspects of magnetic field fluctuations which should 
be incorporated to more realistic F/SOC models of SW driven magnetospheric 
activity.
Scaling and singularity characteristics of high-quality geomagnetic data 
from THULE and NARSSARSSUAQ observatories (Single Observatory Measures) 
and of AE-index (Multi Observatory
 Measure) were compared estimating PDFs, 
flatnesses and singularity rate functions.
 The same methods were applied 
for ACE and WIND data.
 The specific periods chosen for the analysis of SW 
fluctuations reflect 
the limited availability of high resolution satellite 
data for relatively longer 
periods of time. In spite of this, subgroups 
of disturbed and undisturbed periods
 were selected, having primarily in 
mind the occurence of enhanced geomagnetic
 response represented by 1-hour 
mean $D_{st}$-index.
 The comparison of generic features of fluctuations 
in SW and high-latitude SOM and MOM data is then possible, because 
geomagnetic activity at high-latitudes is always very high during 
magnetic storms, though the storm/substorm relationship itself is 
more complicated \cite{dagl01}.

It was shown that  the intermittence of AE-index
 fluctuations is reduced 
at small scales due to the method of its derivation. 
We argue this also 
provides a possible explanation for the negligible changes
 of the AE-index 
singularity distribution with $\tau$. Other kind of MOM data
might be influenced in the same way. Nevertheless, the  contribution of SW
fluctuations make the AE-index rate function asymmetric, mainly within 
the range 
of $\alpha \in (1.2, 1.8)$. The same asymmetry caused by the SW 
is also present
 in THL and NAQ rate functions, however, mainly for 
$\tau > 300$ [min] (see also \cite{voro00}). 
   
In case of SW fluctuations it was demonstrated  that the departure from 
Gaussianity
 is stronger at small scales and it clearly depends on the 
correlation (geoeffectiveness) 
between IMF fluctuations and 
the occurence of geomagnetic storms (decreased
 $D_{st}$-index). It seems 
to indicate that the intermittence strength of IMF magnitude
 and $B_{z}$ 
component fluctuations, 
in addition to other SW parameters such as southward $B_{z}$, SW velocity, density,
Alfvenic Mach number, plasma $\beta$, represents a new parameter (or rather 
a whole
 set of parameters describing singularity features) controlling 
the energy
 input rate to the magnetosphere. Considering Taylor's hypothesis 
and SW velocities of 
500 km/s, intermittence at time scales of tens of 
seconds corresponds to the spatial structures of several thousand kilometers 
or more. \cite{book95} estimated the corresponding timescale on which 
turbulent motion may affect the transport
 of mass and energy across the 
magnetopause through interchange instability. They have found
it is less than 150 [s]. As known from previous ISEE1 and 2
magnetometer studies dayside reconnection of IMF and GMF lines seems often 
to be a 
sporadic and patchy process and measurements obtained at and near 
the magnetopause
 indicate that reconnection does not necessarily occur across 
the all dayside magnetopause
 even under the favourable southward pointing 
IMF conditions \cite{rijn84}. We conjecture, patchy reconnection may be 
related to intermittence and singularity
 characteristics of IMF turbulence 
at small scales.
 There is a number of works in which the role of turbulence 
in magnetopause
 reconnection processes is anticipated 
\cite{gale86, drak94, kuzn95}. 
We recall the work of \cite{gale86} in
which patchy reconnection was considered to be an irregular multiscale 
process associated
 with the magnetic field diffusion and self-consistently 
generated magnetic turbulence.
 Our results indicate that a number of 
singularity parameters (H{\"o}lder exponents) should
 be taken into account 
to properly describe the basic characteristics of the upstream SW
turbulence. In this paper we examined the global distribution of IMF  
singularities and found clear differences between geoeffectively disturbed and 
undisturbed periods. 
Obviously, to understand better the role of turbulence in patchy magnetopause 
reconnection
 processes a proper time and space localization of IMF singularities 
will be needed.

As far as the magnetospheric response is considered, previous results \cite{voro01}
have suggested that  global singularity spectra estimations of SOM 
and MOM data 
sets on different scales may allow to separate fluctuations of SW 
or magnetospheric 
origin. 
Our results show that the influence of the SW is perceptible mainly at the wings of
the rate function, that is at smaller values of $f(\alpha)$. The most probable 
singularities ($f(\alpha) = 0.8 - 1$) are less influenced by SW driver.
Rate functions estimated for years 1975-1996 exhibit similar variations as
geomagnetic activity studied by \cite{kami98}. 
SW forcing effects were found when SOM and MOM singularity spectra for two
years (1984 and 1991) of different 
geomagnetic activity levels were compared.

We believe that further development in this direction will result a better 
understanding of SW - magnetosphere interaction allowing more efficient 
prediction of space weather.

Acknowledgements
The authors wish to acknowledge valuable discussions with Vincenzo Carbone, Giuseppe
Consolini, Manolis Georgoulis, Alex Klimas, Nick Watkins, and Vadim Uritsky.
We are grateful to Yohsuke Kamide and Ioannis Daglis for sending us their results. 
We acknowledge  the use of the Fraclab package developed at the Institut National de 
Recherche en Informatique, Le Chesnay Cedex, France. Geomagnetic data from Thule 
observatory, Narssarssuaq 
observatory and AE-index as well as $D_{st}$-index data 
from WDC Kyoto are gratefully 
acknowledged. We are grateful to N. Ness (Bartol Research Institute) and R. Lepping
(NASA/GSFC) for making the ACE and WIND  data available.
Z. V\"{o}r\"{o}s  and D. Jankovi\v{c}ov\'{a}
were supported by VEGA grant 2/6040.
P. Kov\'{a}cs was supported by the Hungarian Science 
Research Fund (OTKA) under project number F030331 and by the E\"{o}tv\"{o}s 
Scholarship 
provided by the Hungarian Scholarship Committee.

\begin{figure}[tb]
\centerline{
\includegraphics[width=3.2in]{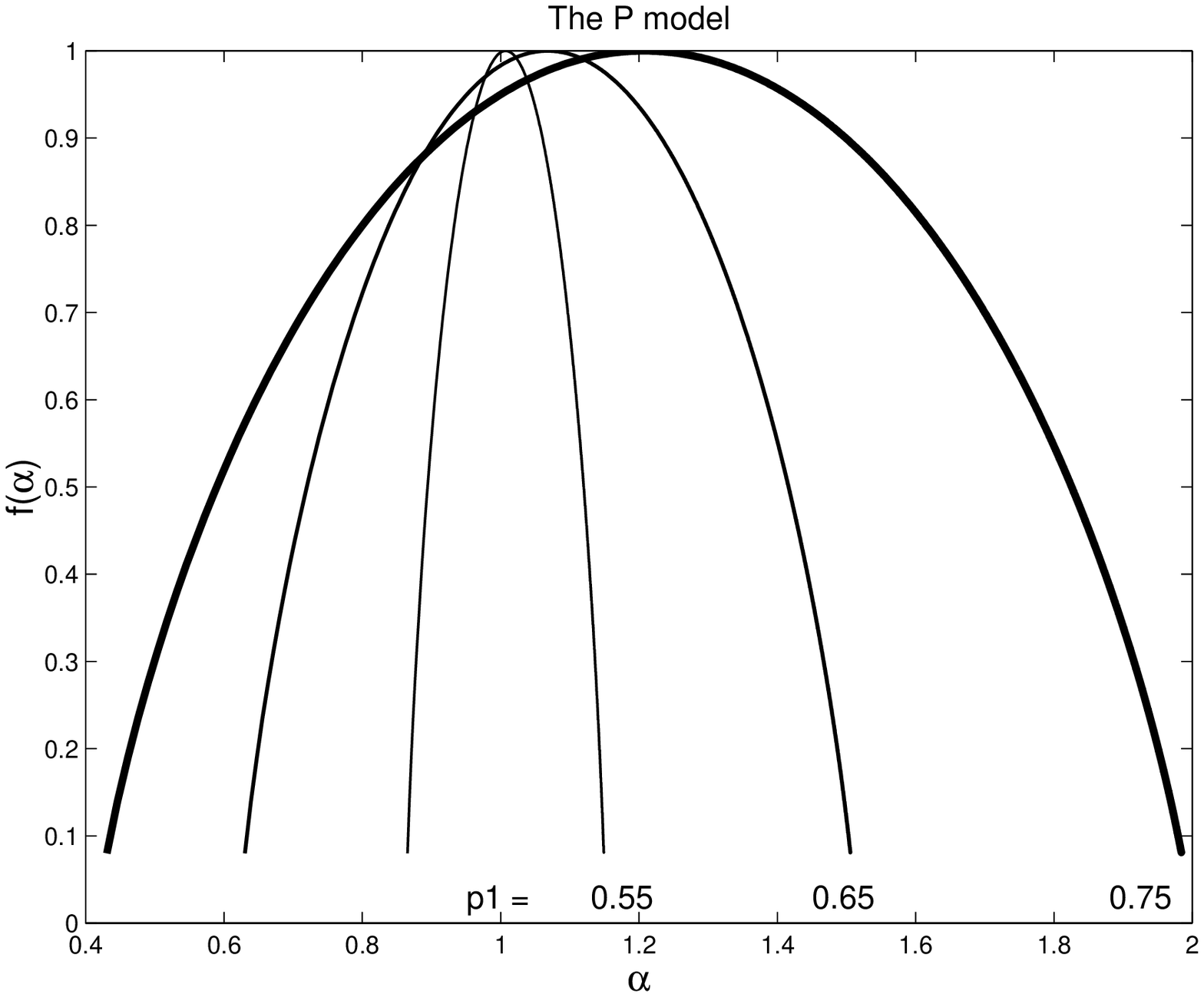}
}
\caption{The $P$-model for three different values of parameter $p_{1}$}
\end{figure}

\begin{figure}[tb]
\centerline{
\includegraphics[width=3.2in]{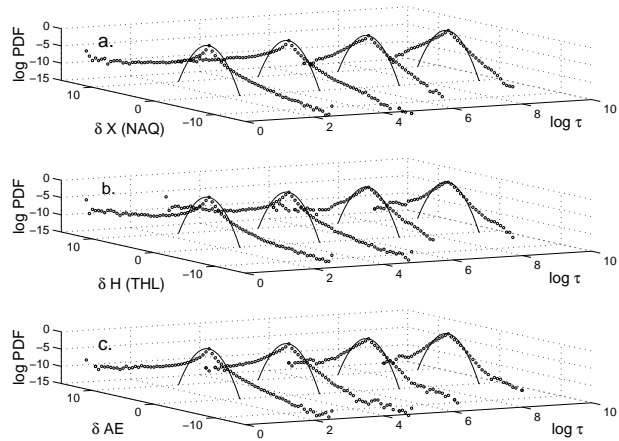}
}
\caption{Probability distribution functions computed for: {\bf a.} NARSSARSSUAQ 
     $X$-component geomagnetic 
field; {\bf b.} THULE $H$-component geomagnetic 
field; 
     {\bf c.} AE-index data. The time scales are $\tau = 5, 50, 500, 5000$ [min].
     Continuous lines refer to Gaussian distribution, while the "o" markers correspond to 
     the analysed data.}
\end{figure}

\begin{figure}[tb]
\centerline{
\includegraphics[width=3.2in]{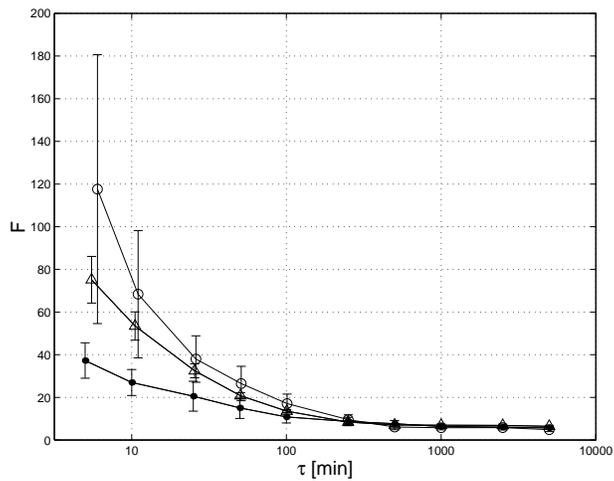}
}
\caption{The flatness as a function of scale ($\tau \in (5, 5000)$ [min]) estimated 
     for NARSSARSSUAQ ($\bigtriangleup$), THULE
 data (o) and AE-index ($\bullet$).}
\end{figure}

\begin{figure}[tb]
\centerline{
\includegraphics[width=3.2in]{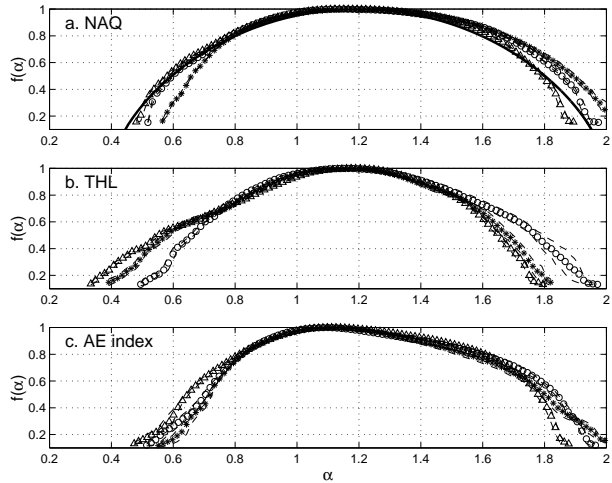}
}
\caption{Dependence of singularity rate functions on scale; $\tau = 5 (\bigtriangleup);
     50 (\ast); 500 (o)$ [min]. {\bf a.} NARSSARSSUAQ data, {\bf b.} THULE data; 
     {\bf c.} AE-index. Thick line in Figure 4a corresponds to $P$-model fit.}
\end{figure}

\begin{figure}[tb]
\centerline{
\includegraphics[width=3.2in]{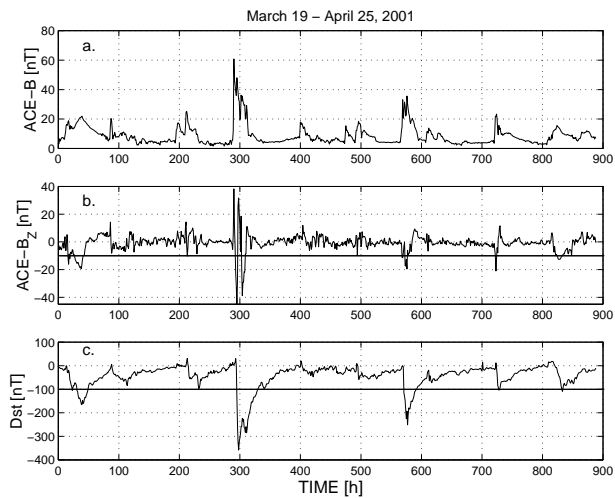}
}
\caption{An extremely disturbed period as seen in 1-hour mean {\bf a.} ACE  IMF magnitude,
     {\bf b.} ACE IMF $B_{z}$ component and
 {\bf c.} $D_{st}$-index data plots.}
\end{figure}

\begin{figure}[tb]
\centerline{
\includegraphics[width=3.2in]{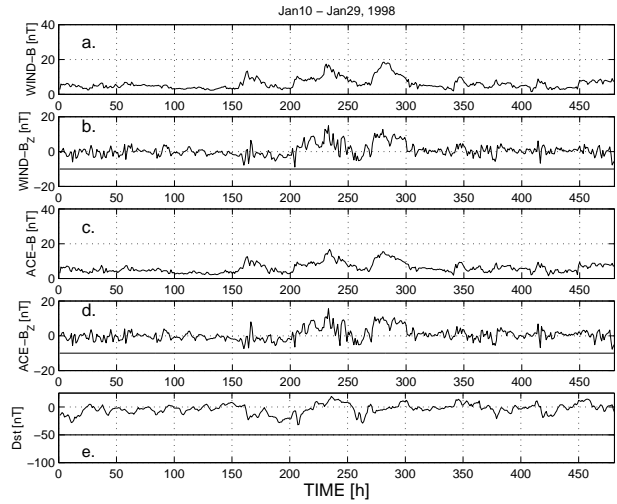}
}
\caption{Undisturbed period as seen in 1-hour mean {\bf a.} IMF $B$ (WIND);
     {\bf b.} IMF $B_{z}$ (WIND); {\bf c.} IMF $B$ (ACE); {\bf d.} IMF $B_{z}$ (ACE) 
     data plots and {\bf e.} the corresponding geomagnetic response represented by 
     $D_{st}$-index.}
\end{figure}

\begin{figure}[tb]
\centerline{
\includegraphics[width=3.2in]{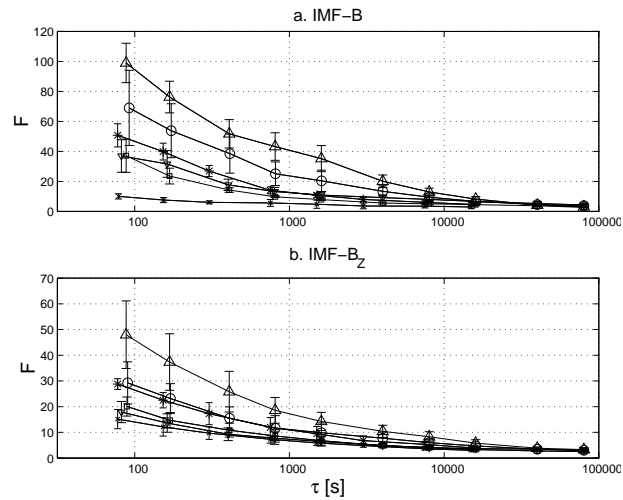}
}
\caption{The flatness as a function of scale ($\tau \in (16*5, 16*10^5)$ [sec]) estimated for 
     the periods: March 19 - April 25, 2001, (ACE: $\bigtriangleup$); 
     October 01 - November 30, 2000, (ACE: o); April 09 - April 20, 1997, (WIND: $\ast$);
     November 18 - December 10, 1998, (ACE: $\bigtriangledown$); January 10 - January 29,
     1998, (WIND: x); January 10 - January 29, 1998, (ACE: $\diamond$). Disturbed, that is
     geoeffective events are depicted by larger markersizes.
 {\bf a.} IMF $B$, 
     {\bf b.} IMF $B_{z}$.}
\end{figure}

\begin{figure}[tb]
\centerline{
\includegraphics[width=3.2in]{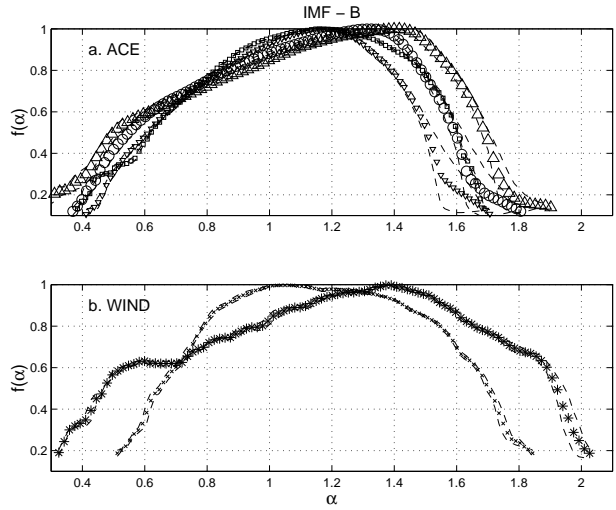}
}
\caption{Singularity spectra (rate functions) estimated for IMF $B$. The periods  
	and data sets are depicted by the same marker types as in Figure 7.}
\end{figure}

\begin{figure}[tb]
\centerline{
\includegraphics[width=3.2in]{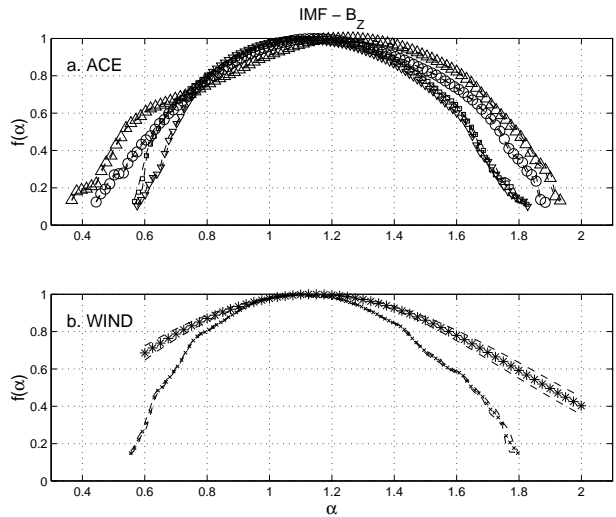}
}
\caption{Singularity spectra estimated for IMF $B_{z}$. The periods and data 
     sets are depicted by the same marker types as in Figure 7.}
\end{figure}

\begin{figure}[tb]
\centerline{
\includegraphics[width=3.2in]{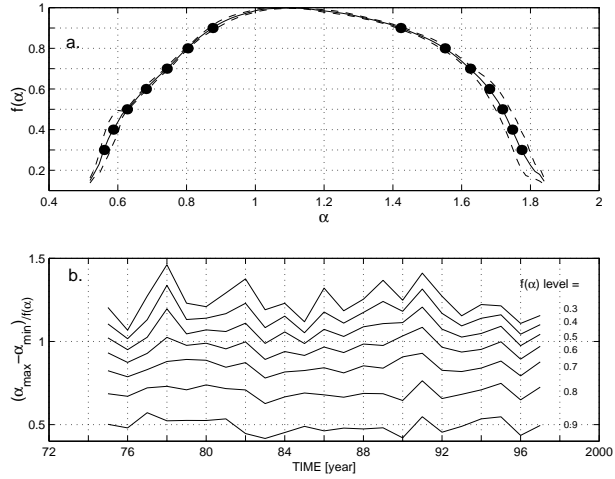}
}
\caption{Time variation of the shape of singularity spectra {\bf a.} filled 
     circles show the points in which $\alpha_{min}$ and $\alpha_{max}$ 
     values are computed; {\bf b.} $\alpha_{max} - \alpha_{min}$ values 
     computed at different $f(\alpha)$ levels.}
\end{figure}

\begin{figure}[tb]
\centerline{
\includegraphics[width=3.2in]{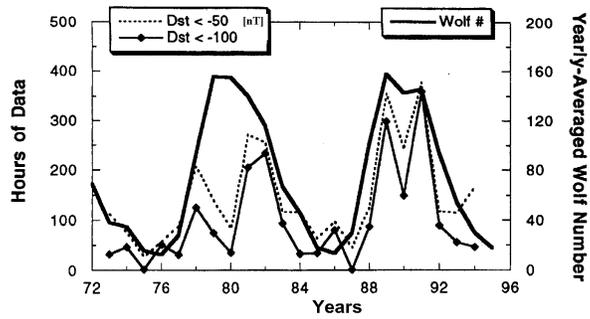}
}
\caption{Yearly averaged number of hours with $D_{st}$ $<$ -100 nT 
     ($\diamond$), and with $D_{st}$ $<$ -50 nT ($- \: - \: -$).
     Yearly averaged sunspot number (thick line) is also shown. After \cite{kami98}.}
\end{figure}

\begin{figure}[tb]
\centerline{
\includegraphics[width=3.2in]{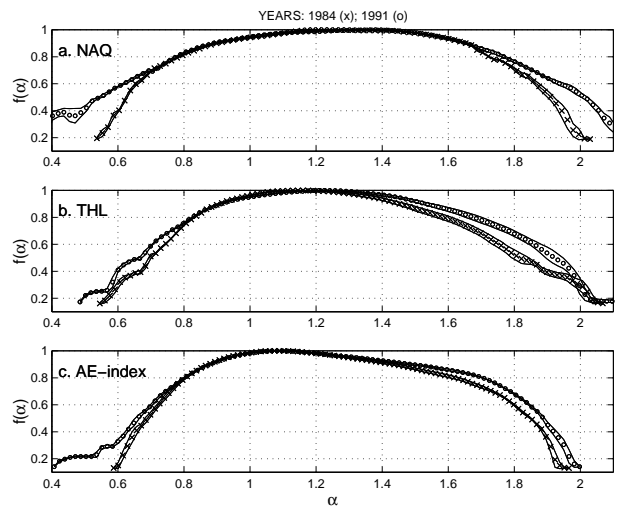}
}
\caption{Singularity spectra for {\bf a.} NARSSARSSUAQ data; {\bf b.} 
     THULE data; {\bf c.} AE-index. Maximum of geomagnetic activity 
     in 1991 (o); minimum of geomagnetic activity in 1984 (x).}
\end{figure}

\end{document}